\documentstyle[epsbox,12pt]{article} 
\normalbaselines
\begin{document}
\bibliographystyle{unsrt}
\begin{center}
{\large \bf Some properties of quantum reiliablity function for 
quantum communication channel}\\[7mm]
K.~Kurokawa, $^\ast$M.~Sasaki, M.~Osaki, and O.~Hirota\\[5mm]
{\it Research Center for Quantum Communications, \\
Tamagawa University, Tokyo 194 (Japan)}\\[5mm]
{\it $^\ast$ Communication\ Research\ Laboratory,\\
Ministry\ of\ Posts\ and\ Telecommunications}\\[5mm]
\end{center}

\vspace{2mm}

\begin{abstract}
This paper presents  some examples of quantum reliability function for the
 quantum communication system  in which  classical information is transmitted by 
 quantum states. In addition, the quantum Cut off rate is defined.  
They will be compared  with  Gallager's reliability function for the same system. 
\end{abstract}

\section{Introduction}
Recently, quantum communication and information theory has received much 
attention, because many distinguished features as quantum nature were discovered 
in comparison with conventional information theory.  It is well known that Shannon's
 first and second theorems  play the most important role in the conventional one, 
 which give an existence of effective coding schemes in information source and 
 channel[1]. Recently, the quantum version of Shannon's first theorem was proved 
 by Schumacher[2].  On the second theorem,  Stratonovich[3] and Holevo[4] pointed 
 out that a true capacity for quantum measurement channel is greater than the 
 maximum mutual information with respect to detection operators and a priori 
 probability for information symbols, and that it will be bounded by von Neumann 
 entropy for an ensemble of signal states. Hausladen et. al[5] proved quantum  
 version of Shannon's second theorem by introducing a typical sub-space in 
 addition to random coding scheme and square root measurement.  
 This corresponds to a typical sequence in the conventional theory.  
 As a result, it was shown that the zero error  channel capacity is von Neumann 
 entropy for an ensemble of signal states.
On the other hand, we remain  problems how to derive the maximum mutual 
information with respect to detection operators and a priori probability for 
information symbols so called un-coded capacity : 
${C}_{\rm 1}$
.  In this problem, Holevo's 
necessary condition formula[6] and Davies's theorem[7] play an important role. 
Fuchs et. al [8]tried to give some examples of ${C}_{\rm 1}$, 
and Ban[9] and Osaki[10] proved
  that if the signal states are group covariant, then the square root measurement 
  and minimax solution in the detection theory satisfy Holevo's  
  necessary condition, and showed some examples of ${C}_{\rm 1}$. 
  Also Sasaki[11] 
  recently gave some example for super additivity of capacity and its coding 
  scheme in order to connect a gap between ${C}_{\rm 1}$ and von Neumann entropy as 
  the true channel 
  capacity. 
As a natural extension, Holevo[12] presented a theory of quantum reliability 
function corresponding to Gallager's reliability function in the conventional theory[13].
 The theory of reliability function is very important in 
 communication engineering. So it is also essential to clarify  detailed properties 
 of quantum reliability function in order to study coding scheme with finite length.  
In this paper,  we shall show numerical examples of Holevo's quantum reliability 
function for several quantum state signals, and compare with 
Gallager's reliability function to quantum measurement channel.\\
\\
\section{A theory of quantum reliability function}
\subsection{Reliability function}
Here we suvey a theory of reliability function for quantum channel.
Let us consider quantum channel with a finite input pure state:
${\rho }_{i}\rm =\left|{{\mit \psi }_{i}}\right\rangle\left\langle{{\psi }_{i}}\right|$.
According to quantum version of Shannon's second theorem by Hausladen, the zero 
error channel capacity is given by von Neumann entropy as follows:\\
\begin{equation}
H\rm ({\mit S}_{\xi }\rm )=-Tr[{\mit S}_{\xi }\rm ln{\mit S}_{\xi }\rm ]
\label{eqn:von}
\end{equation}
\begin{equation}
{S}_{\xi }\rm =\sum\nolimits\limits_{}^{} {\mit \xi }_{i}{\rho }_{i}
\label{eqn:S}
\end{equation}
The von Neumann entropy for quantum information source is larger than ${C}_{\rm 1}$.
 In order to take into account this fact in reliability function theory, the next 
 formalism can be used.
\begin{equation}
E_Q\rm (\mit R\rm )=\lim_{\mit n\rm \rightarrow \infty }sup
{1 \over \mit n}\rm \ln{1 \over {\mit P}_{e}}
\label{eqn:1}
\end{equation}
where Pe is average error probability of code words, and $R$
 is information transmission  rate.  One cannot caluculate 
${E}_{Q}\rm (\mit R\rm )$
 directly. However, if  there is  a bound for this ${E}_{Q}\rm (\mit R\rm )$ ,  
 then one can estimate the following inequality.
 \begin{equation}
{P}_{e}\rm \le {\mit e}^{\rm -\mit n{E}_{Q}\rm (\mit R\rm )}
\le {\mit e}^{\rm -\mit n{E}_{Qr}\rm (\mit R\rm )}
\label{eqn:2}
\end{equation}
An upper bound for the average error probability for code words was derived based 
 on square root measurement and random coding technique by Holevo[12].\\
\\
{\large $<$Theorem$>$}\\

For any $M$ and $\rm 0\le \mit s\rm \le 1$
, the following upper bound is valid:
\begin{equation}
{P}_{e}\rm \le 2{(\mit M\rm -1)}^{\mit s}{\left[{\rm Tr{\mit S}_
{\xi }^{\rm 1+\mit s}}\right]}^{\mit n}
\label{eqn:3}
\end{equation}
In the above theorem,  we can set \\
\begin{center}
$M\rm ={\mit e}^{\mit nR}\rm >1$
：code word$.$\\
${S}_{\xi }$：density operator$.$\\
\end{center}
\begin{equation}
{\left[{\rm Tr{\mit S}_{\xi }^{\rm 1+\mit s}}\right]}^{\mit n}\rm =
{\mit e}^{\mit n\rm \ln Tr{\mit S}_{\xi }^{\rm 1+\mit s}}
\label{eqn:4}
\end{equation}
When one inserts the above relations into Eq(\ref{eqn:3}), we have 
\begin{equation}
{P}_{e}\rm \le 2{\mit e}^{\mit nsR}\rm \times {\mit e}^{\mit n\rm \ln Tr{\mit S}_
{\xi }^{\rm 1+\mit s}}\rm =2{\mit e}^{\mit n\rm \{\ln Tr{\mit S}_
{\xi }^{\rm 1+\mit s}\rm +\mit sR\rm \}}
\label{eqn:5}
\end{equation}
Let us rewrite the above equation as follows:
\begin{equation}
{P}_{e}\rm \le 2{\mit e}^{\rm -\mit n\rm \{\mit \mu \rm (\mit s\rm ,\mit \xi \rm )-
\mit sR\rm \}}
\label{eqn:6}
\end{equation}
where 
\begin{equation}
\mu \rm (\mit s\rm ,\mit \xi \rm )=-\l nTr{\mit S}_{
\xi }^{\rm 1+\mit s}
\label{eqn:u}
\end{equation}
As a result, we can define
\begin{equation}
{E}_{Qr}\rm (\mit R\rm )\equiv \max_{\mit s}\max_{\xi}
\left[\mu \rm (\mit s\rm ,\mit \xi \rm )-\mit sR\right]
\label{eqn:7}
\end{equation}
The maximization with respect to "$s$"  is given by
\begin{equation}
{\rm \partial  \over \partial \mit s}\rm \left[{{\mit \mu }^{}\rm (\mit s\rm ,\mit \xi \rm )-
\mit sR}\right]\rm ={\partial \left[{{\mit \mu }^{}\rm (\mit s\rm ,\mit \xi \rm )}\right] 
\over \partial \mit s}\rm -\mit R\rm =0
\label{eqn:8}
\end{equation}
where we have 
\begin{equation}
{\rm \partial \mit \mu \rm (\mit s\rm ,\mit \xi \rm ) \over \partial \mit s}\rm =
{-Tr{\mit S}_{\xi }^{\rm 1+\mit s}\rm \ln{\mit S}_{\xi } \over \rm Tr{\mit S}_
{\xi }^{\rm 1+\mit s}}\rm ={-\sum\nolimits\limits_{}^{} {\mit \lambda }_
{\mit j}^{\rm 1+\mit s}\rm \ln{\mit \lambda }_
{\mit j} \over \rm \sum\nolimits\limits_{}^{} {\mit \lambda }_
{\mit j}^{\rm 1+\mit s}}
\label{eqn:9}
\end{equation}
For the maximization with respect to a priori probability, it is natural way.\\
\\
\subsection{Cut off rate}
In the conventional theory, we define sometimes Cut off rate which is the special 
case of reliability function [14]. Here we can define also quantum version of 
Cut off rate based on Eq(10) as follows:
\begin{equation}
{R}_{\rm 0}\equiv \max\limits_{\mit \xi }^{}\mu \left({{\xi }_{i}\rm ,\mit s\rm =1}\right)
=\max-\ln{\rm Tr}{\mit S}_{\xi }^{\rm 2}
=\max-\ln\sum\nolimits\limits_{}^{} \sum\nolimits\limits_{}^{} {\mit \xi }_{i}{\xi }_{j}
{\left|{\left\langle{{\psi }_{i}}\right|\left.{{\psi }_{j}}\right\rangle}\right|}^{\rm 2}
\label{eqn:R0}
\end{equation}
This is exactly a maximization of the entropy introduced by Stratonovich [3].      
The maximization  is very simple. That is, one needs only to optimize a priori probability as same with classical one.  
In the quantum reliability function, let us assume that the signal power is enough large, 
which corresponds to almost orthogonal states.   In this case, the optimization with respect
 to "s" does not make sense. So the quantum  reliability function in the case with large photon
  signals may be replaced by the quantum Cut off rate. \\
In future problems, when we encounter more difficult problems  in which  calculation of 
reliability function like in the conventional theory is so difficult, the quantum cut off 
rate will provide useful way.
\section{Examples for several quantum state signals.}
Let us here calculate quantum version of reliability function for several quantum signals. 
 For our purpose, the next theorem is useful.\\
 \\
{\large $<$Theorem$>$}[3]\\

The eigenvalues of density operator for ensemble of  signal states are equal to 
those of the Gram matrix for signal set.\\
\\
(A)  Binary, 3 ary PSK and 4 ary orthogonal signal by pure state\\

Let us first assume that the signal states are $\left|{\alpha }\right\rangle\rm ,
\left|{-\mit \alpha }\right\rangle$.
The Gram matrix for this signal is given by\\
\[\left[
\begin{array}{cc}
\sqrt {{\xi }_{\rm 1}}\left\langle{\mit \alpha }\right.\left|{\alpha }\right\rangle\sqrt {{\xi }_
{\rm 1}}
&\sqrt {{\xi }_{\rm 1}}\left\langle{\mit \alpha }\right.\left|{\rm -\mit \alpha }\right\rangle
\sqrt {{\xi }_{\rm 2}}
\\
\sqrt {{\xi }_{\rm 2}}\left\langle{-\mit \alpha }\right.\left|{\alpha }\right\rangle\sqrt {{\xi }_
{\rm 1}}
&\sqrt {{\xi }_{\rm 2}}\left\langle{-\mit \alpha }\right.\left|{\rm -\mit \alpha }\right
\rangle\sqrt {{\xi }_{\rm 2}}
\end{array}
\right]\]
\begin{equation}
=\left[
\begin{array}{cc}
{\xi }_{\rm 1}
&\kappa \rm \sqrt {{\mit \xi }_{\rm 1}{\mit \xi }_{\rm 2}}
\\
\kappa \rm \sqrt {{\mit \xi }_{\rm 1}{\mit \xi }_{\rm 2}}
&{\xi }_{\rm 2}
\end{array}
\right]
\label{eqn:10}
\end{equation}
where\\
\begin{center}
$\kappa \rm =\left\langle{\mit \alpha }\right.\left|{\rm -\mit \alpha }\right\rangle\rm =
exp\left[{-2{\left|{\mit \alpha }\right|}^{\rm 2}}\right].$\\
\end{center}
The eigenvalues of the above matrix are  
\begin{equation}
{\lambda }_{\rm 1}={1 \over 2}
\left[{1-\sqrt {1-4(1-{\mit \kappa }^{\rm 2})
\mit \xi \rm (1-\mit \xi \rm )}}\right]
\label{eqn:r1}
\end{equation}
\begin{equation}
{\lambda }_{\rm 2}={1 \over 2}\left[{1+
\sqrt {1-4(1-{\mit \kappa }^{\rm 2})\mit \xi \rm 
(1-\mit \xi \rm )}}\right]
\label{eqn:r2}
\end{equation}   
As a result, we have  the next relation.\\
\begin{equation}
{\mu }_{r}\rm (\mit \xi \rm )
{\rm \equiv }
-\mit \ln\left\{{{\lambda }_{\rm 1}^{1+\mit s}\rm 
+{\mit \lambda }_{\rm 2}^{1+\mit s}}\right\}\rm 
-\mit sR\rm =\mit \mu \rm (\mit s\rm ,\mit \xi \rm )
-\mit sR
\label{eqn:11}
\end{equation}
\begin{equation}
{\rm \partial \mit \mu \rm (\mit s\rm ,\mit \xi \rm ) 
\over \partial \mit s}\rm = {-{\mit \lambda }_
{\rm 1}^{1+\mit s}\rm \ln{\mit \lambda }_{\rm 1}
-{\mit \lambda }_{\rm 2}^{1+\mit s}
\rm \ln{\mit \lambda }_{\rm 2} 
\over {\mit \lambda }_{\rm 1}^{1+\mit s}
\rm +{\mit \lambda }_{\rm 2}^{1+\mit s}}\rm 
-\mit R\rm =0
\label{eqn:12}
\end{equation}
From Eqs (16) and (17), we can find quantum reliability function for this signals.
The figure 1-a  shows a numerical example. Then, based on the same procedure mentioned 
above, we can give the quantum reliability functions for 3 ary PSK and 
$M$
 ary orthogonal 
signal systems. Figure 1-b and 1-c show the numerical examples for them. 
The signal states of 3-aryPSK and orthogonal signal are as follows:
\begin{equation}
\left|{\alpha }\right\rangle\rm ,
\left|{\mit \alpha {e}^{i{\rm 2 \over 3}
\mit \pi }}\right\rangle\rm ,
\left|{\mit \alpha {e}^{\rm -\mit i
{\rm 2 \over 3}\mit \pi }}\right\rangle
\label{eqn:PSK}
\end{equation}
\begin{equation}
\matrix{\left|{{\psi }_{\rm 1}}\right\rangle=
{\left|{\mit \alpha }\right\rangle}_{\rm 1}
{\left|{0}\right\rangle}_{2}{\left|{0}\right\rangle}_{3}
{\left|{0}\right\rangle}_{4}\cr
\left|{{\mit \psi }_{\rm 2}}\right\rangle=
{\left|{0}\right\rangle}_{1}
{\left|{\mit \alpha }\right\rangle}_{\rm 2}
{\left|{0}\right\rangle}_{3}
{\left|{0}\right\rangle}_{4}\cr
\left|{{\mit \psi }_{\rm 3}}\right\rangle=
{\left|{0}\right\rangle}_{1}{\left|{0}\right\rangle}_{2}
{\left|{\mit \alpha }\right\rangle}_{\rm 3}
{\left|{0}\right\rangle}_{4}\cr
\left|{{\mit \psi }_{\rm 4}}\right\rangle=
{\left|{0}\right\rangle}_{1}
{\left|{0}\right\rangle}_{2}
{\left|{0}\right\rangle}_{3}
{\left|{\mit \alpha }\right\rangle}_{\rm 4}\cr}
\end{equation}
\\
In these figures,  
solid line means the part of quantum reliability function with optimum value 
$s\rm =1$. 
Dashed line means that with optimum value $s\rm <1$.\\
\\
(B) Ternary pure states signal\\

Here let us consider ternary pure states given as 
$\left|{\rm 0}\right\rangle,
\left|{\mit \alpha }\right\rangle\rm ,
\left|{-\mit \alpha }\right\rangle$
Ternary signal is non-symmetric and the optimization with respect to a priori 
probability is necessary[15].  Hence, in this case, 
 we should clarify the maximum von Neumann entropy, because the maximum 
value is not given by equal a priori probability, while equal a priori probability gives the 
maximum von Neumann entorpy in cases of binary, PSK, and orthogonal signal systems. 
The Gram matrix for this signal is given by
\begin{equation}
\left[
\begin{array}{ccc}
{\xi }_{\rm 1}
&\kappa \rm \sqrt {{\mit \xi }_{\rm 1}{\mit \xi }_{\rm 2}}
&\kappa \rm \sqrt {{\mit \xi }_{\rm 1}{\mit \xi }_{\rm 3}}
\\
\kappa \rm \sqrt {{\mit \xi }_{\rm 2}{\mit \xi }_{\rm 1}}
&{\xi }_{\rm 2}
&\kappa^4 \rm \sqrt {{\mit \xi }_{\rm 2}{\mit \xi }_{\rm 3}}
\\
\kappa \rm \sqrt {{\mit \xi }_{\rm 3}{\mit \xi }_{\rm 1}}
&\kappa^4 \rm \sqrt {{\mit \xi }_{\rm 3}{\mit \xi }_{\rm 2}}
&{\xi }_{\rm 3}
\end{array}
\right]
\label{eqn:13}
\end{equation}
\begin{center}
$\kappa \rm ={\mit e}^{\rm -{\mit Ns \over \rm 2}}$\\
${N}_{s}\rm ={\left|{\mit \alpha }\right|}^{\rm 2}$
：signal photon$.$\\
\end{center}
Let  ${\lambda }_{\rm 1},{\mit \lambda }_{\rm 2},{\mit \lambda }_{\rm 3}$
  be eigenvalues of the Gram matrix. Then the von Neumann entorpy becomes
\begin{equation}
\mit H\rm ({\mit S}_{\xi }\rm )=-{\mit \lambda }_{\rm 1}
\ln{\mit \lambda }_{\rm 1}-{\mit \lambda }_{\rm 2}
\ln{\mit \lambda }_{\rm 2}-{\mit \lambda }_{\rm 3}
\ln{\mit \lambda }_{\rm 3}
\label{eqn:14}
\end{equation}
The von Neumann entropy as function of a priori probabilities can be visualized as 
shown in figures  2-a and 2-b.   Thus, a priori probabilities which give the maximum 
von Neumann entropy are not uniform. By taking into account this fact, we show the 
quantum reliability function for ternary signals in figure 3.\\
\\
\section{ Application of Gallager's reliability function.}
Here we apply Gallager's reliability function to quantum channel. In general it is defined 
as follows:
\begin{equation}
E\left(R\right)=\max_{\rho,\xi}E'
\label{eqn:15}
\end{equation}
\begin{equation}
E'=-\rho R-\ln\sum_{j=1}^k 
\left(\sum_{i=1}^r \xi_i{\mit p}_{ij}^{1/1+\rho}\right)^
{1+\rho}
\label{eqn:16}
\end{equation}
\begin{center}
$0<\rho\leq1,$\quad $0\leq\xi_i\leq1,$\\
$\xi=\{\xi_1,\cdots,\xi_r\}$：a priori probability,\\
$k$：number of output symbol, \\
$r$：number of input symbol.\\
\end{center}
In order to apply Gallager's reliability function to quantum system, we first have to 
define channel matrix . As an example, we here treat the binary case. In this case, 
we employ the optimum detection operators for average error probability.  So the 
channel matrix can be represented as follows:
\begin{equation}
\left[
\begin{array}{cc}
\frac{2F+\lambda-2\kappa^2+1}{4F}&
\frac{2F-\lambda+2\kappa^2-1}{4F}\\
\frac{2F-\lambda+2\lambda\kappa^2-1}{4F}&
\frac{2F+\lambda-2\lambda\kappa^2+1}{4F}\\
\end{array}
\right]
\label{eqn:17}
\end{equation}
where\\
\begin{center}
$\displaystyle F=
\sqrt{\left(\frac{1+\lambda}{2}\right)^2-\lambda\kappa^2}$,
 \quad 
$\lambda=\xi_1/\xi_2$,\\
$\kappa=\left<\alpha|-\alpha\right>=
\exp\left[-2|\alpha|^2\right]$.\\
\end{center}
From Eqs(\ref{eqn:15}), (\ref{eqn:16}),  
we can find Gallager's reliability function. In figure 4, we 
show the numerical property in comparison with the quantum case. As a result, we 
obtaine that the zero error capacity in case of Gallager is ${C}_{\rm 1}$, and it 
cannot achieve 
von Neumann entropy.\\
\section{ Conclusions}
In this paper,  we surveyed a theory of quantum reliability function and defined quantum 
Cut off  rate. Then we examined quantum reliability function for several quantum signals.  
As a result,  if signal power is enough large, the quantum reliability function is almost 
same as quantum cut off rate. In addition , it was shown that Gallager's functions for 
quantum systems provide only a coding scheme based on un-coded capacity ${C}_{\rm 1}$.
\\

\begin{center}
\epsfile{file=sk2,height=4cm,width=8cm,scale=1}
\end{center}
\begin{center}
\epsfile{file=skp3,height=4cm,width=8cm,scale=1}
\end{center}
\begin{center}
\epsfile{file=skt4,height=4cm,width=8cm,scale=1}
\end{center}
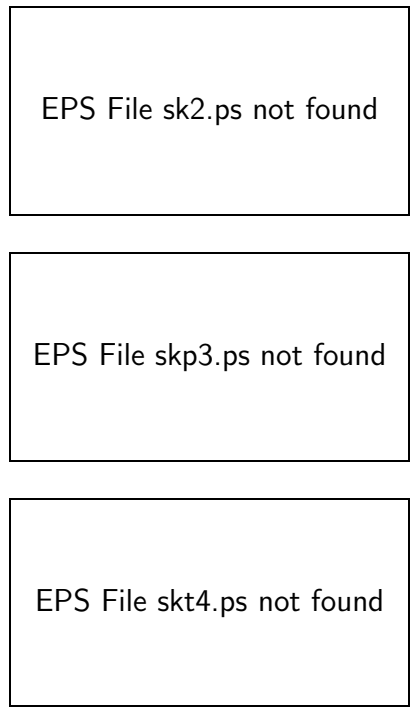
Fig.1 Numerical properties of quantum reliability function for binary(1-a),
3-ary PSK(1-b), and 4-ary orthogonal signal(1-c).\\
\newpage
\begin{center}
\epsfile{file=g1,height=5cm,width=5cm,scale=1}
\epsfile{file=g2,height=4.5cm,width=4.5cm,scale=1}
\end{center} 
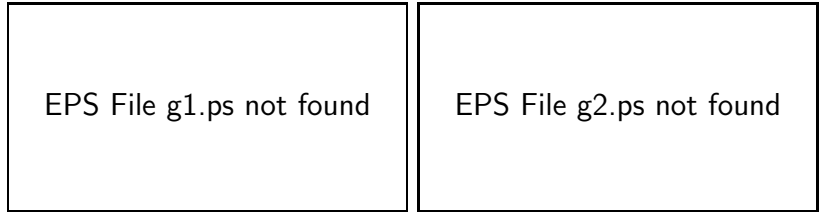
Fig.2 Numerical property of von Neumann entropy.\\ 
\begin{center}
\epsfile{file=sk3,height=4cm,width=8cm,scale=1}
\end{center} 
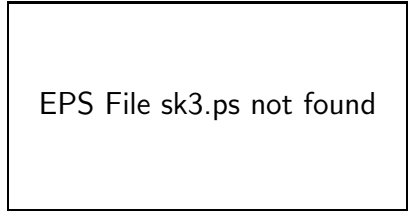
Fig.3 Quantum reliability function for ternary.\\ 
\begin{center}
\epsfile{file=g4,height=4cm,width=8cm,scale=1}
\end{center} 
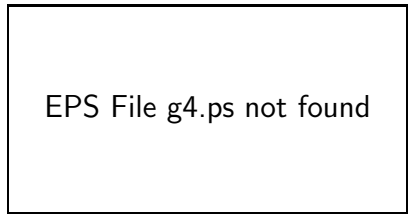
Fig.4 Comparison with quantum and Gallager reliability functions.\\
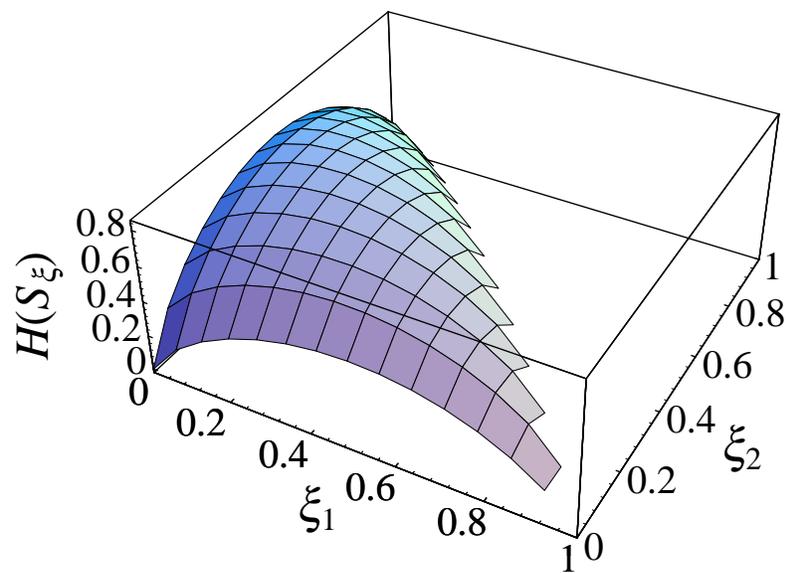
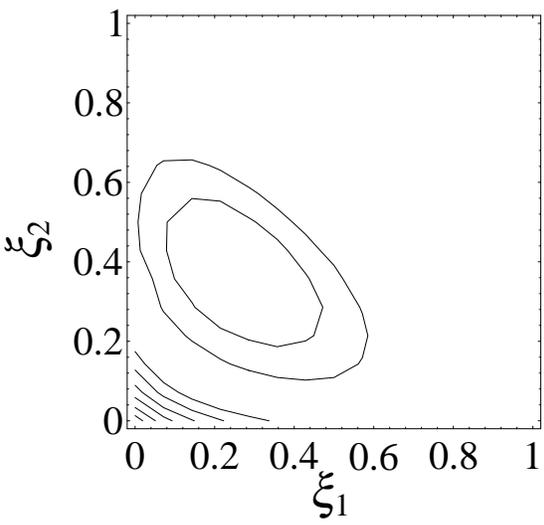
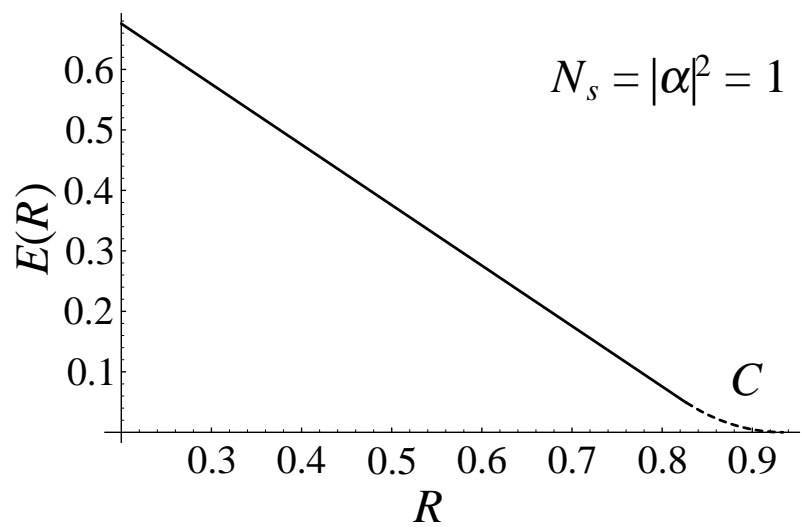
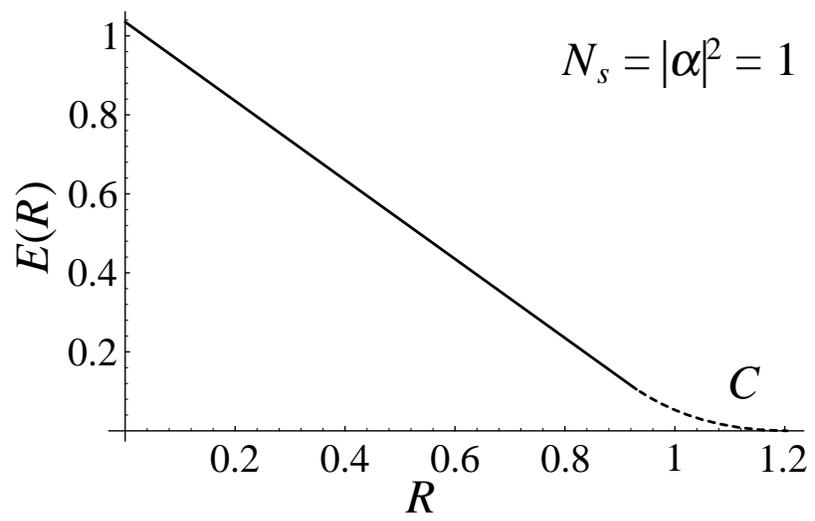
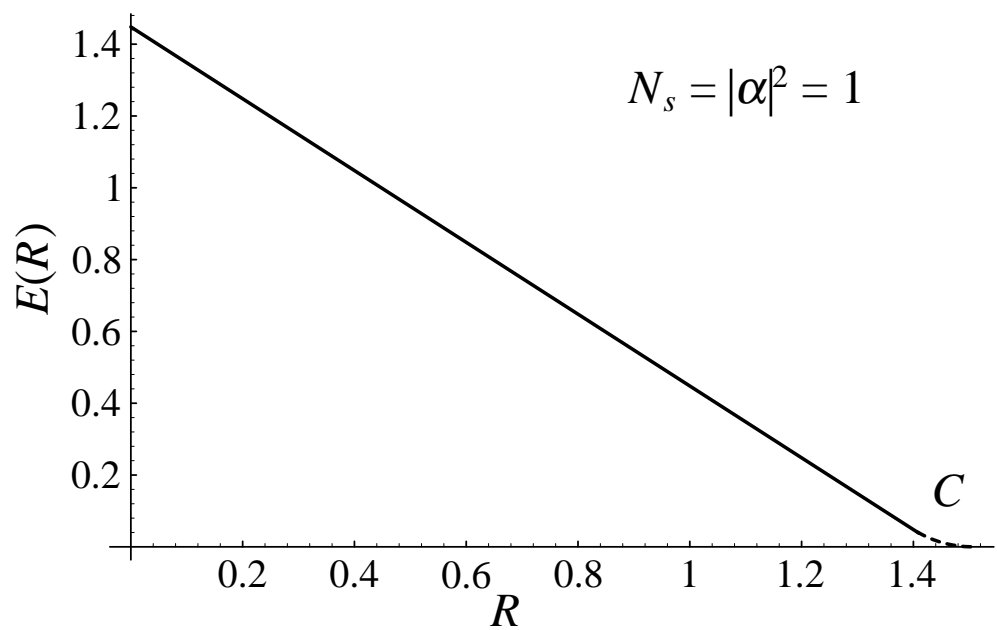
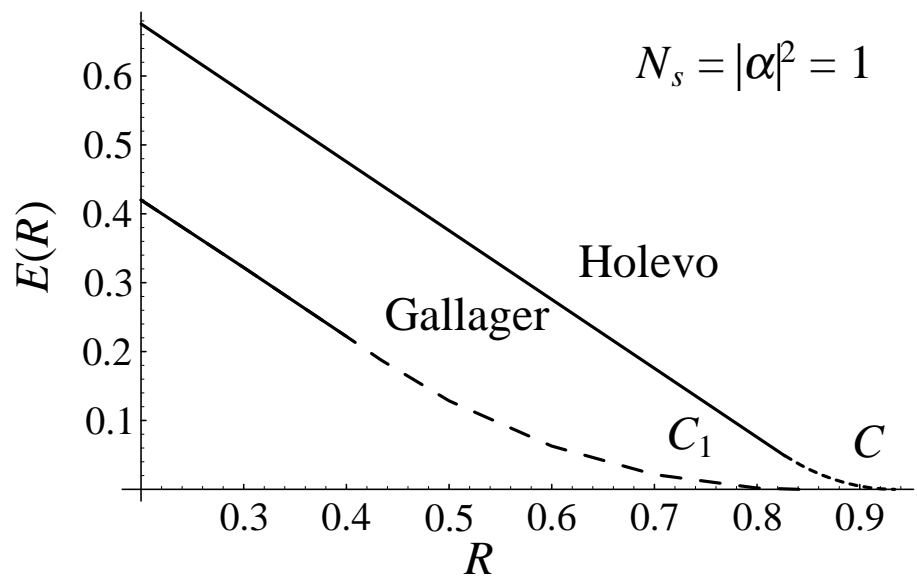
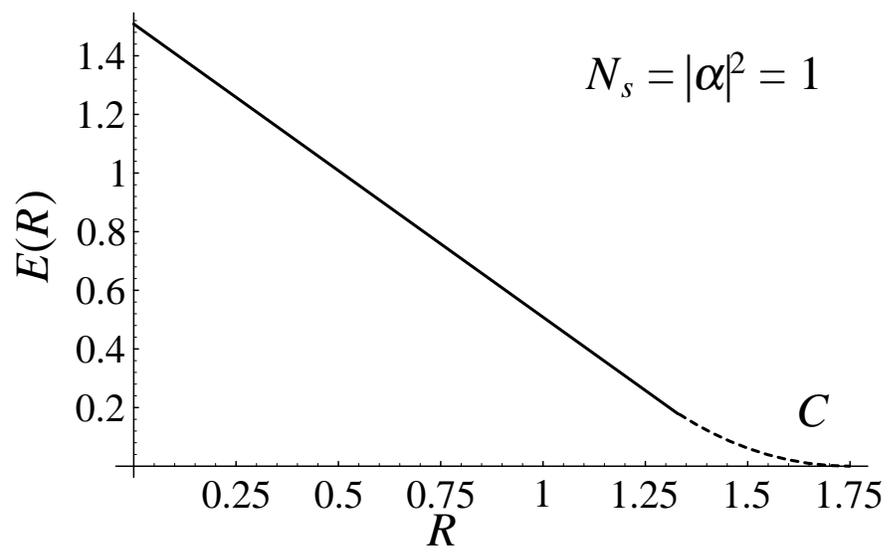
\end{document}